# An optical diode made from a 'flying' photonic crystal

Da-Wei Wang [1, 2 *], Jörg Evers [1, 3] and Shi-Yao Zhu [1, 2, 4]

[1] *Beijing Computational Science Research Center, Beijing 100084, China*
[2] *Centre of Optics Sciences and Department of Physics, The Chinese University of Hong Kong, Hong Kong, China*
[3] *Max-Planck-Institut für Kernphysik, Saupfercheckweg 1, 69117 Heidelberg, Germany*
[4] *The State Key Laboratory of Quantum Optics and Quantum Optics Devices, Institute of Opto-Electronics, Shanxi University, Taiyuan 030006, China*

**Optical diodes controlling the flow of light are of principal significance for optical information processing [1]. They transmit light from an input to an output, but not in reverse direction. This breaking of time reversal symmetry is typically achieved via non-linear [2,3] or magnetic effects [4], which imposes limits to all-optical control [5-7], on-chip integration [7-11], or single-photon operation [12]. Here, we propose an optical diode which requires neither magnetic fields nor strong input fields. It is based on a flying photonic crystal. Due to the Doppler effect, the crystal has a band gap with frequency depending on the light propagation direction relative to the crystal motion. Counter-intuitively, our setup does not involve the movement of any material parts. Rather, the flying photonic crystal is realized by optically inducing a spatially periodic but moving modulation of the optical properties of a near-resonant medium. The flying crystal not only opens perspectives for optical diodes operating at low light levels or integrated in small solid state devices, but also enables novel photonic devices such as optically tunable mirrors and cavities.**

Optical diodes are also known as optical isolators [4], as they are used to isolate lasers from reflections from the actual experiment to improve their stability. The class of media suitable for diode operation is constrained by the Lorentz reciprocal theorem, which loosely speaking states that under certain conditions, source and detection position for light passing through a device can be interchanged without modifying the transmission properties. The magnetic Faraday effect, which breaks the time-reversal symmetry of a polarization rotation [4], and non-symmetrically acting nonlinearities [2,3,13-15] can circumvent this constraint by violating its requirements. But these approaches are at odds with desirable features such as the ability to integrate the diode on a chip [1,7,8,10], to make it optically controllable [5-7], and to reduce its operation range to low field intensities [12]. For example, Faraday rotators are typically large and magnetic fields cause unwanted effects in semiconductors devices, while nonlinearities require large electric fields of the probe light. Recently, also alternative implementations of optical diodes based on photonic interband transitions [16], liquid crystals [17] and optomechanical cavity [18] have been proposed.

The key element of the optical diode implementation proposed here is a photonic crystal generated from a periodic modulation of the optical properties of a medium, which leads to the formation of a band gap. Light is transmitted through the crystal if its frequency is outside the gap, but it is not transmitted if the frequency is inside the gap. A static photonic crystal has the same transmission properties for light of equal frequency incident from opposite directions. But if the photonic crystal moves, then the Doppler effect blue shifts light counter-propagating to the motion, but red-shifts co-propagating light. If in the reference frame of the crystal one of the shifted frequencies is inside the gap, while the other is outside, then only light in one of the two propagation directions is transmitted, and a diode is formed.

However, this naive picture raises the question, how a flying photonic crytal could be realized in practice. In the following, we will show how this is possible without moving any material parts. Instead of using a conventional photonic crystal based on a periodic sequence of different materials, we use an optically controllable photonic crystal based on an electromagnetically induced transparency (EIT) system driven by a standing wave control laser field [19-21]. The standing wave pattern formed by two counter-propagating coupling fields imposes a periodic modulation of the refractive index, which acts as a photonic crystal for a weak probe field. By detuning the two counter-propagating components of the standing wave against each other, the envelope of the standing wave and thus the optically induced photonic crystal can be moved in a static medium. It is therefore the standing wave pattern that moves rather than the medium itself.

We next turn to the details of the implementation. We consider a medium forming a three level system as shown in Figure 1 (a). The energy difference between states $|i\rangle$ and $|j\rangle$ ($i, j = a, b, c$) is $\omega_{ij}$, and the transition between $|a\rangle$ and $|c\rangle$ is forbidden by symmetry. The states $|a\rangle$ and $|b\rangle$ are coupled by a detuned standing wave $\mathbf{E}_c(t) = \hat{y}[E_1 \cos(\omega_1 t - k_1 x) + E_2 \cos(\omega_2 t + k_2 x)]$ where $E_1$ and $E_2$ are the amplitudes of the two counter-propagating fields with frequency $\omega_1$ and $\omega_2$. $k_i = \omega_i / c$ is the magnitude of the wave vectors of field $E_i$. With $\delta = \omega_2 - \omega_1$ and $\omega_c = (\omega_1 + \omega_2)/2$, the velocity of the flying standing wave envelope is

$$v = -\frac{\delta}{2\omega_c} c. \tag{1}$$

For $\delta \neq 0$, a sign reversal of the detuning $\delta \to -\delta$ changes the sign of the velocity, $v \to -v$, and the envelope flies in the opposite direction. Equation (1) is the origin of the time-reversal symmetry breaking required for the diode operation [22,23]. As $\delta$ and $t$ only enter the analysis as a product, a time reversal $t \to -t$ can equivalently be regarded as a detuning reversal $\delta \to -\delta$. Due to the Doppler shift, the transmission for nonzero $\delta$ and $-\delta$ will be different as the medium properties are frequency dependent, and thus the time reversal symmetry is broken.

A weak forward probe field $\mathbf{E}_f(t) = \hat{z} E_f \cos(\omega_f t - k_f x)$ probing the transition between $|c\rangle$ and $|b\rangle$ can be reflected by the moving photonic crystal. Due to the Doppler effect, the reflected field $\mathbf{E}_b(t) = \hat{z} E_b \cos(\omega_b t + k_b x)$ has frequency $\omega_b = \omega_f + \delta$. The probe and reflected fields in the flying photonic crystal can be described with multi-wave mixing method efficiently. The coupled wave equations for probe and reflected fields are

$$\frac{\partial}{\partial x} E_f(x) = -\beta_{12}(\omega_f) E_f(x) + i\kappa_{21}(\omega_b) e^{-i\Delta k_x x} E_b(x), \tag{2}$$

$$-\frac{\partial}{\partial x}E_b(x) = -\beta_{21}(\omega_b)E_b(x) + i\kappa_{12}(\omega_f)e^{i\Delta k_x x}E_f(x), \qquad (3)$$

where $\beta_{ij}(\omega)$ ($\kappa_{ij}(\omega)$) is the attenuation (wave-mixing) coefficient for the probe field of frequency $\omega$ and propagating along with the coupling field $E_i$ and anti-parallel to $E_j$. The coefficients are related to the linear and nonlinear medium susceptibilities $\chi_{ij}^{L}$ and $\chi_{ij}^{NL}$ by $\beta_{ij} = k_p \cos\theta \operatorname{Im}\chi_{ij}^{L}/2$ and $\kappa_{ij} = k_p \cos\theta E_1 E_2 \chi_{ij}^{NL}/2$ where $\theta$ is the angle between the probe and the coupling fields, as shown in Figure 1 (b). $\Delta k_x = k_p \cos\theta \left[ 2 + \left(\operatorname{Re}\chi_{12}^{L} + \operatorname{Re}\chi_{21}^{L}\right)/2 \right] - 2k_c$ is the wave vector mismatch along the $\hat{x}$-direction. In calculating $\chi_{ij}^{L}$ and $\chi_{ij}^{NL}$, we take into account all orders of the coupling fields, $E_1$ and $E_2$ for the generation of the reflected field frequency, e.g., via four-wave ($\omega_b = \omega_f - \omega_1 + \omega_2$), six-wave ($\omega_b = \omega_f - \omega_1 + \omega_2 - \omega_1 + \omega_1$), and higher-order mixing. The summation of all orders of the coupling fields converges and can be performed by the technique of continued fractions [24,25]. For a sample of length $L$, the reflectance and transmittance are obtained as

$$R \equiv \left|\frac{E_b(0)}{E_f(0)}\right|^2 = \left| \frac{1}{\kappa_{21}(\omega_b)} \frac{e^{-(\lambda_+ - \lambda_-)L} - 1}{e^{-(\lambda_+ - \lambda_-)L}\left[\lambda_+ + \beta_{12}(\omega_f)\right]^{-1} - \left[\lambda_- + \beta_{12}(\omega_f)\right]^{-1}} \right|^2, \qquad (4)$$

$$T \equiv \left|\frac{E_f(L)}{E_f(0)}\right|^2 = \left| \frac{e^{(\lambda_+ + \lambda_-)L}(\lambda_- - \lambda_+)}{\left[\lambda_- + \beta_{12}(\omega_f)\right]e^{\lambda_- L} - \left[\lambda_+ + \beta_{12}(\omega_f)\right]e^{\lambda_+ L}} \right|^2, \qquad (5)$$

where

$$\lambda_{\pm} = \frac{-i\Delta k_x - \beta_{12}(\omega_f) + \beta_{21}(\omega_b)}{2} \pm \frac{\sqrt{\left[i\Delta k_x - \beta_{12}(\omega_f) - \beta_{21}(\omega_b)\right]^2 + 4\kappa_{12}(\omega_f)\kappa_{21}(\omega_b)}}{2}. \qquad (6)$$

The non-reciprocity from the perspective of wave-mixing lies in the frequencies of the two coupled fields. As shown in Figure 1 (b), for a left-to-right incident probe field the two coupled waves have frequencies $\omega_f$ and $\omega_b = \omega_f + \delta$, while for a right-to-left incident probe field the two coupled waves have frequencies $\omega_f$ and $\omega_b = \omega_f - \delta$. As both $\beta_{ij}(\omega)$ and $\kappa_{ij}(\omega)$ are frequency dependent, the reflectance and transmittance are different in these two cases.

In Figure 2 (a) and (b), the reflectance and transmittance $R_L$ and $T_L$ ($R_R$ and $T_R$) for the light incident from the left (right) side are plotted as functions of the probe detuning $\Delta_p \equiv \omega_{bc} - \omega_p$. As an example, we use the parameters for three states of the Caesium $D_1$ line, $6^2S_{1/2}, F=4$ ($|a\rangle$), $6^2P_{1/2}, F=4$ ($|b\rangle$) and $6^2S_{1/2}, F=3$ ($|c\rangle$). We fix the one-photon detuning of $E_1$ to $\Delta_1 \equiv \omega_{ba} - \omega_1 = 0$ and tune $\delta$ via $\omega_2$. To further characterize the diode performance, in Figure 2 (c), the contrast of the two transmittances $\eta = (T_L - T_R)/(T_L + T_R)$ is plotted for the two optical diodes flying with opposite velocities in Figure 2 (a) and (b). For $\eta = 1$ ($\eta = -1$), a diode transmitting only from left to right (right to left) is achieved. As expected from the Doppler effect, the band gaps experienced by the left- and right- incident probe fields are shifted with respect to each other by the coupling field detuning $\delta$. From Figure 2 (a), we find two working regions for the optical diode with opposite pass direction, depending on the probe frequency. Around $\Delta_p = -0.1\gamma_{bc}$ (or $\Delta_p = 0.1\gamma_{bc}$), where $\gamma_{bc}$ is the decoherence rate between levels $|b\rangle$ and $|c\rangle$, an optical diode is obtained that only allows transmission from left to right (right to left) with the contrast ratio of more than 200dB/mm. As shown in Figure 2 (b), by changing the sign of $\delta$ (reversing the flight direction), the two working regions interchange their relative positions.

Figure 3 shows the transmission spectra in the two opposite directions as a function of the standing wave component detuning $\delta$. The purple areas (labelled as BG) between the two transmission branches are the band gaps. It can be seen that the pass bands of the left incident light correspond to band gaps of the right incident light and vice versa over a broad range of detunings $\delta \neq 0$. We also note that our flying crystal has properties different from those of a truly moving material photonic crystal. First, the band gap width increases linearly with the detuning, as opposed to the "real" photonic crystal, for which the motion does not change the band gap width. Another difference to a real flying photonic crystal is $E_1 \neq E_2$, so that the profiles of the two transmission spectra cannot be overlapped by a simple displacement of $\delta$.

At fixed $\delta$, only the two-photon detuning $\Delta_p - \Delta_1$ is relevant in the spectra whose profile is not sensitive to the one-photon detuning of the coupling field $\Delta_1 \equiv \omega_{ba} - \omega_1$ as long as $\Delta_1$ is not much larger than the Rabi frequency of the coupling fields (a few times larger is still fine). We can therefore tune the working point of the diode easily by changing $\Delta_1$ while keeping $\delta$ fixed. The ratio of the Rabi frequencies of the two standing wave components is another crucial parameter. If $(E_2/E_1)^2 \approx 1$, the overall transmission is small because of high absorption due to vanishing EIT at the standing wave nodes [19]. For $(E_2/E_1)^2 \approx 0$, the band gap is narrow and the usual EIT transmission is obtained. For the parameters in Figure 2, we found that the ratio $(E_2/E_1)^2 = 0.2$ is a good compromise. The results are essentially insensitive to a variation of $\theta$ in a range of $1°$, because at the chosen density the dispersion outweighs the angle-induced phase mismatch by two orders.

Depending on the actual implementation, the medium density and length can be limited. We found that our results approximately depend on the product of density and sample length $NL$. Thus, a reduction in density can be compensated by a longer sample and vice versa. In the following, we exploit this feature to propose a proof-of-principle implementation of our diode scheme based on room temperature EIT experiments with atoms [20]. Due to the atom motion the coefficients in Eqs. (2) and (3) are Doppler broadened. Compared to Figs. 2 and 3, the atom density is lowered to $N = 4 \times 10^{10} \text{cm}^3$, compensated by a sample length increase to 2 cm. Results are shown in Figure 4. Although the transmittance in pass direction is reduced to about 50% (red line at $\Delta_p = 2\gamma_{bc}$), the more crucial extinction of the signal in the stop direction still can exceed 30dB/cm, ensuring an efficient isolation.

EIT is a hallmark effect of quantum optics, well studied both in theory and in experiments [15], and has been integrated on chips [26,27] and reduced to the single photon level [28]. This together with the flexible and robust scaling of our method to different parameter regimes opens perspectives for flying photonic crystal diodes operating at low light levels across a broad range of media, e.g., integrated in small solid state devices.

Going beyond the concept of a diode, a flying photonic crystal with high reflectance as in Figure 2 can also be exploited to construct novel photonic devices. For example, the flying crystals can serve as the walls of a cavity which allow photons to enter but prevents them from leaving, or visa-versa. The properties of this cavity can dynamically be controlled via the 'velocities' of the walls, e.g., to trap or release photons on demand. Furthermore, due to the "velocity"-dependent Doppler shift upon reflection [see Figure 1 (b)], the probe photon frequency can be modulated by dynamically changing the "velocities" of the walls.

In conclusion, we proposed an optical diode based on a flying photonic crystal. The diode is non-magnetic and linear in the probe field response, such that it is suitable for probe fields of low intensities. The flying crystal is induced optically without actually moving any material parts. The diode can be optically controlled and implemented in a broad parameter range. We found contrast ratios up to 200dB/mm at density $4 \times 10^{14} cm^3$ which could be obtained in solid state media, and also proposed a proof-of-principle implementation based on state-of-the-art room temperature EIT with contrast ratios exceeding 30dB/cm. The flying crystal concept also enables novel photonic devices such as optically tunable mirrors and cavities.

We thank RB Liu for helpful discussion. We thank A. Crosse for the proof reading and the insightful suggestions on physics. This work was supported by National Basic Research Program of China (2011CB922203) and National Natural Science Foundation of China (11174026).

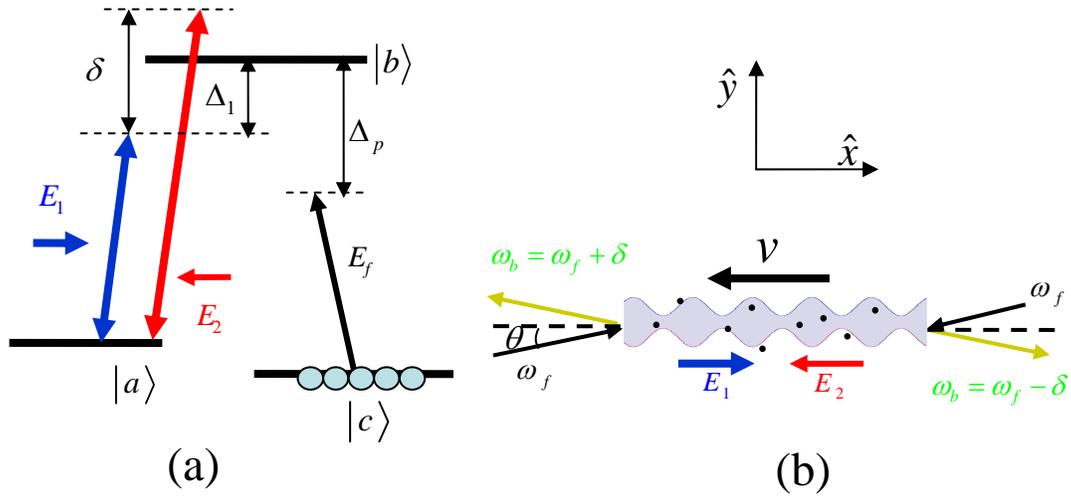

**Figure 1. Implementation of the flying photonic crystal diode.** (a) Level structure of the EIT medium. The coupling transition of the $\Lambda$-type three-level system is driven by a standing wave field with detuning $\delta$ between the two components. (b) The photonic crystal is formed by the standing wave pattern which flies to the left if $\delta > 0$. The black dots indicate the static medium atoms with level-structure as in (a). The photonic band gaps for light incident parallel or anti-parallel to the crystal movement are shifted with respect to each other in frequency due to the Doppler effect and an optical diode is achieved.

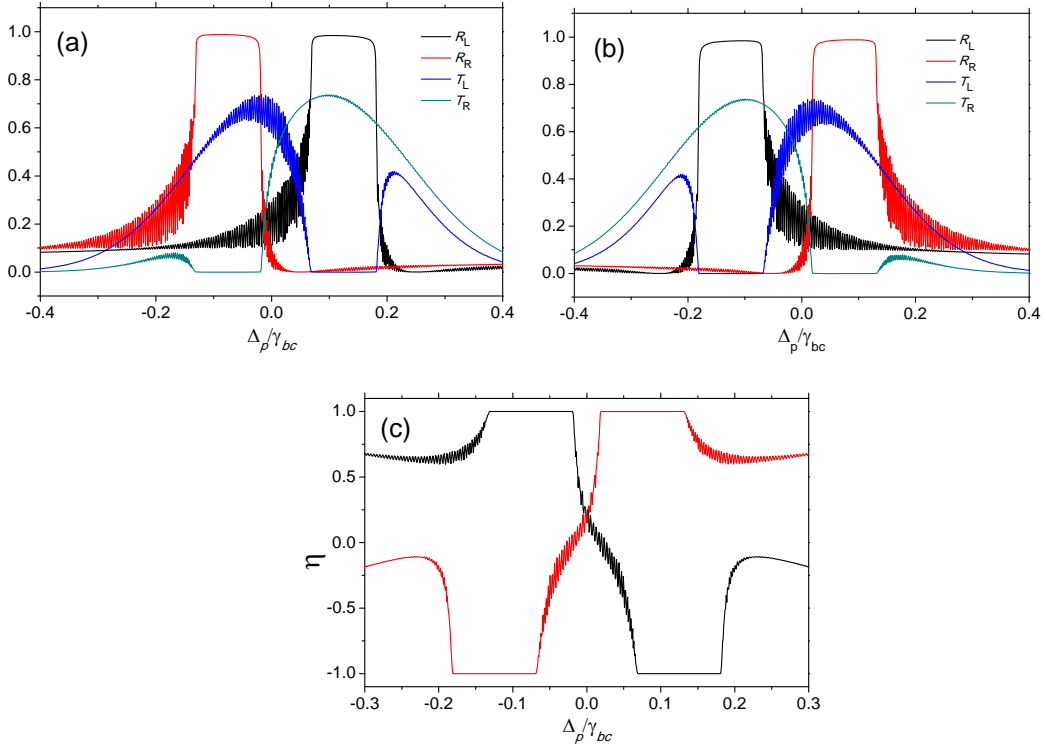

**Figure 2. Reflection and transmission characteristics of the diode**. The reflectances $R_L$ (black) and $R_R$ (red) and transmittances $T_L$ (blue) and $T_R$ (green) of the optical diode for both left (subscript 'L') and right (subscript 'R') incident light. The standing wave detuning is (a) $\delta = 0.2\gamma_{bc}$, and (b) $\delta = -0.2\gamma_{bc}$. The detuning of $E_1$ is zero, $\omega_1 = \omega_{ba}$. In (c), the contrast of the transmittance $\eta$ for (a) (black) and (b) (red) is plotted. The Rabi frequency of the coupling field $E_1$ is $10\gamma_{bc}$ and $(E_2/E_1)^2 = 0.2$. The density of the atoms is $N = 4 \times 10^{14} \text{cm}^3$. The length of the atomic sample is $L = 1\text{mm}$. The incident angle is $\theta = 0$. The dephasing rate between the two ground states is chosen as zero.

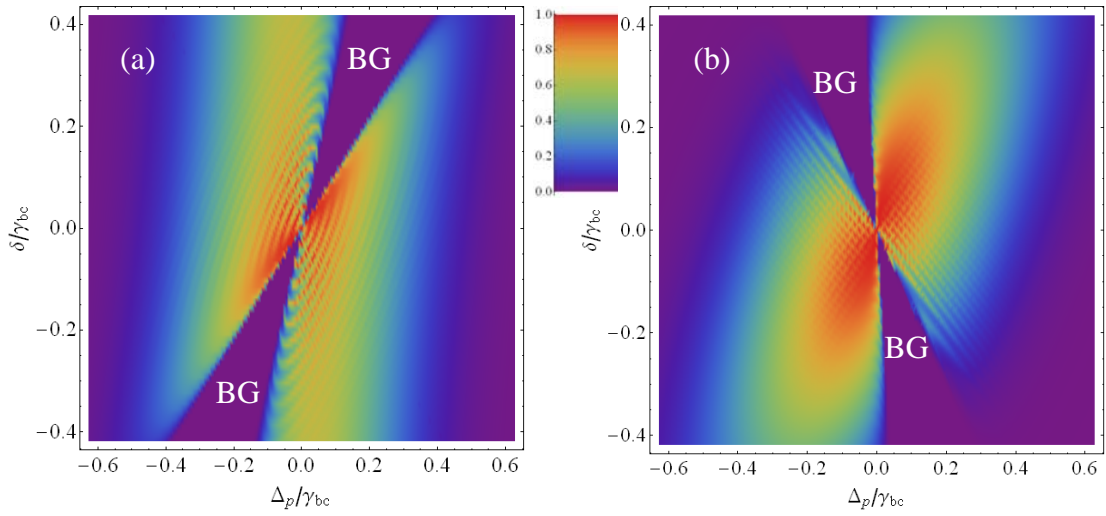

**Figure 3. Band gap structure as a function of the detunings.** The transmission spectra $T_L$ in (a) and $T_R$ in (b) are plotted as a function of the detuning $\delta$. 'BG' denotes the band gaps. The band gaps in (a) overlap the pass bands in (b) and vice versa. The parameters are the same as in Figure 2.

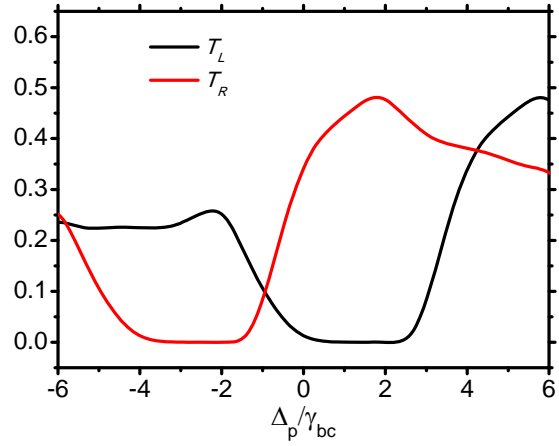

**Figure 4. Possible proof-of-principle implementation in a thermal atom vapor.** The figure shows the optical diode performance in a Doppler broadened atomic gas at temperature 300K. Parameters are $E_1 = E_2$ with Rabi frequency $10\gamma_{bc}$, $\Delta_1 = 2\gamma_{bc}$, $\delta = 4\gamma_{bc}$, ground state decoherence rate $\gamma_{ac} = 0.1\gamma_{bc}$, sample length 2cm, atom density $N = 4\times 10^{10}\,\text{cm}^3$, and incidence angle $\theta = 0$.

\* e-mail: cuhkwdw@gmail.com